\documentclass[pra,twocolumn,showpacs,amsmath,amssymb]{revtex4}
\usepackage{graphicx,bm,color}
\newtheorem{theorem}{Theorem}
\usepackage[utf8]{inputenc}

\begin{document}

\title{Central frequency of few-cycle laser pulses in strong-field processes}

\author{J. Venzke}\thanks{J.V. and T.J. contributed equally to this work}
\author{T. Joyce}\thanks{J.V. and T.J. contributed equally to this work}
\author{Z. Xue}
\author{A. Becker}
\author{A. Jaron-Becker}
\affiliation{
 JILA and Department of Physics, University of Colorado, Boulder, CO 80309-0440, USA
}
\begin{abstract}
We analyze the role of the difference between the central frequencies of the spectral distributions of the vector potential and the electric field of a short laser pulse. The frequency shift arises when the electric field is determined as the derivative of the vector potential to ensure that both quantities vanish at the beginning and end of the pulse. We derive an analytical estimate of the frequency shift and show how it affects various light induced processes, such as excitation, ionization and high harmonic generation. Since observables depend on the frequency spectrum of the electric field, the shift should be taken into account when setting the central frequency of the vector potential to avoid potential misinterpretation of numerical results for processes induced by few-cycle pulses.
\end{abstract}


\date{May 22, 2018}

\maketitle

\section{Introduction}

Few-cycle laser pulses are used in many interesting strong-field applications (for reviews, see e.g., \cite{popmintchev10,suzuki14,vrakking14,pazourek15,wang15,xu16}): For example, high-order harmonics and (isolated) attosecond pulses are generated, ultrafast atomic and molecular dynamics as well as charge transfer and exciton dynamics can be induced and time resolved, molecular structure can be imaged on ultrashort time scales, or chemical reactions may be controlled. Therefore, light sources generating ultrashort intense laser pulses in different regions of the spectrum, at extreme ultraviolet \cite{goulielmakis08}, ultraviolet \cite{durfee99}, optical \cite{fork87}, near-infrared \cite{zhou94,nisoli96,nisoli97} and infrared wavelengths \cite{ishii14,pupeza15,li16}, have been developed over the past decades.  The simulation of the time-dependent response of matter to a few-cycle pulse, e.g. via the numerical solution of the corresponding time-dependent Schr\"odinger equation (TDSE), can however crucially depend on the definition of the electric field $E(t)$ used. To achieve quantitative agreement between theory and experiment, the potential issues present in both numerical calculations and experiment must be well understood and minimized.

As pointed out by Chelkowski and Bandrauk \cite{chelkowski02}, the representation of $E(t)$ via an envelope function times a trigonometric function may lead to a non-vanishing potential $A(t)$ at the end of the pulse. This inconsistency can be resolved by first defining the magnitude of the vector potential $A(t)$ as (we use Hartree atomic units: $e = m_e = \hbar =1$) \cite{chelkowski02}: 
\begin{equation}
A(t) = f(t) \cos(\omega_A t +\phi_A),
\label{eq:vectorpotential}
\end{equation}
where $f$, $\omega_A$, and $\phi_A$ are the envelope function, central frequency, and carrier envelope phase of the vector potential, respectively. The amplitude of the electric field is then obtained via the derivative:
\begin{equation}
\begin{split}
\label{eq:efield}
E(t) =& -\frac{1}{c}\frac{\partial}{\partial t}A(t)
\\
=& -\frac{\omega_A}{c} f(t) \sin(\omega_A t +\phi_A) 
\\
&- \frac{1}{c}\frac{\partial f(t)}{\partial t}
\cos(\omega_A t +\phi_A).
\end{split}
\end{equation}
With this definition it is guaranteed \cite{chelkowski02} that both vector potential and electric field vanish at the beginning and the end of the pulse.

The expression for the electric field, Eq.~(\ref{eq:efield}), includes a term that depends on the time derivative of $f$ and, hence, may have significant effects in the case of few-cycle pulses. As we will show below, a direct implication is that the central frequency of the electric field spectrum, $\omega_E$, is not equal to $\omega_A$. The frequency shift $|\omega_E - \omega_A|$ is small for long pulses but it increases for a decrease of the pulse duration. Consequently, numerical results obtained for linear processes---such as excitation and ionization---involving few-cycle pulses with the same value for the central frequencies $\omega_A$ and $\omega_E$ do not coincide. The frequency shift is also noticeable for nonlinear processes, such as two-photon excitation and high-order harmonic generation, and it scales with the number of photons involved. Since observable quantities depend on the frequency spectrum of the electric field, the frequency shift should be taken into account when setting the central frequency $\omega_A$ of the vector potential in a calculation.

The paper is organized as follows: In Section \ref{sec:pulse}, we present an analytical estimate for the magnitude of the frequency shift, and discuss how to correct for it.
In Section \ref{sec:applications}, we present results of numerical solutions of the time-dependent Schrödinger equation, which illustrate how the frequency shift affects a number of quantum mechanical processes: photoionization, resonant excitation, and high-harmonic generation. We end with a brief summary.

\section{Estimation of Frequency Shift}
\label{sec:pulse}

\begin{figure}[t]
\centering
   
   \includegraphics[width=0.49\linewidth]{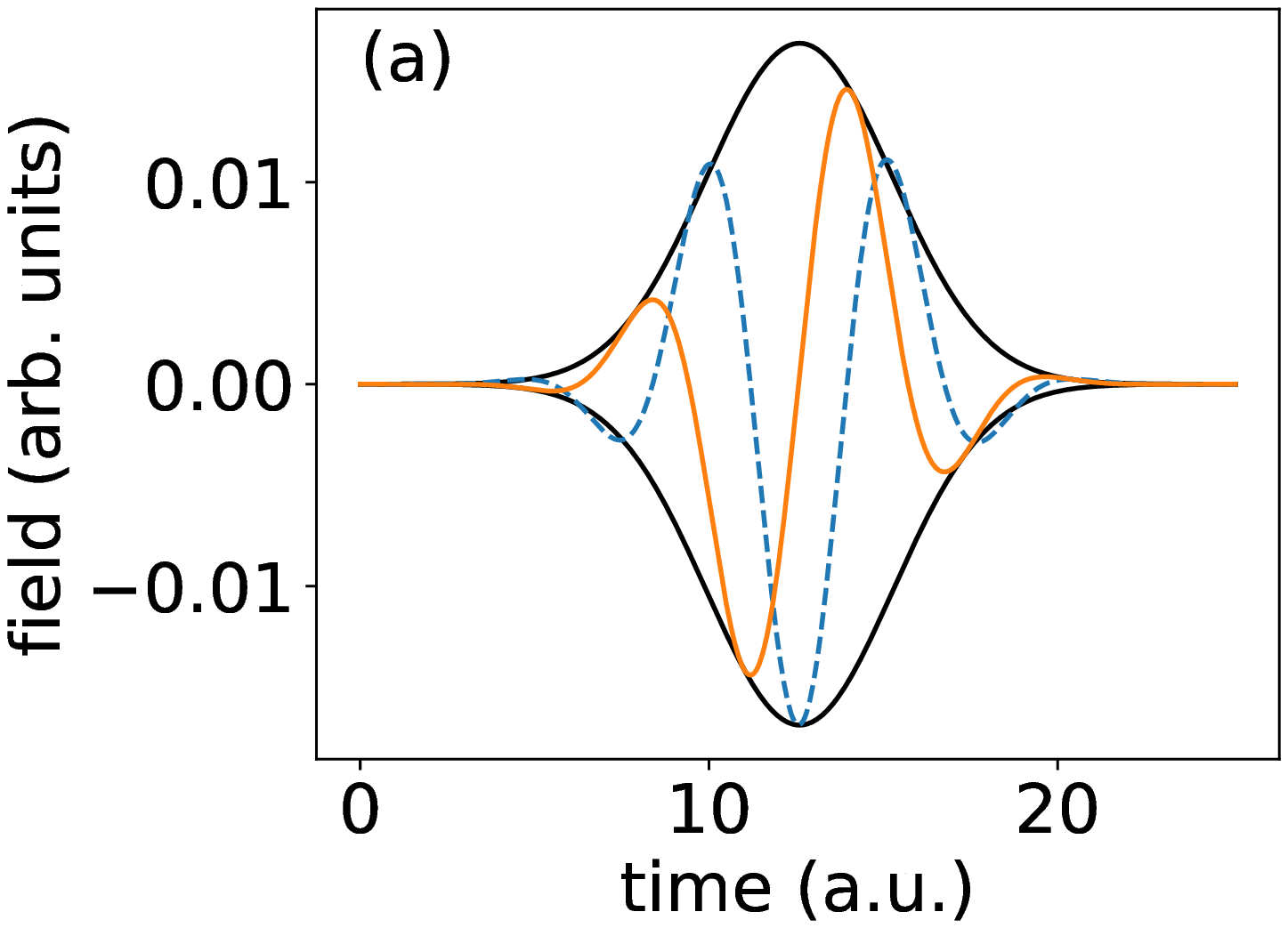}
   \includegraphics[width=0.49\linewidth]{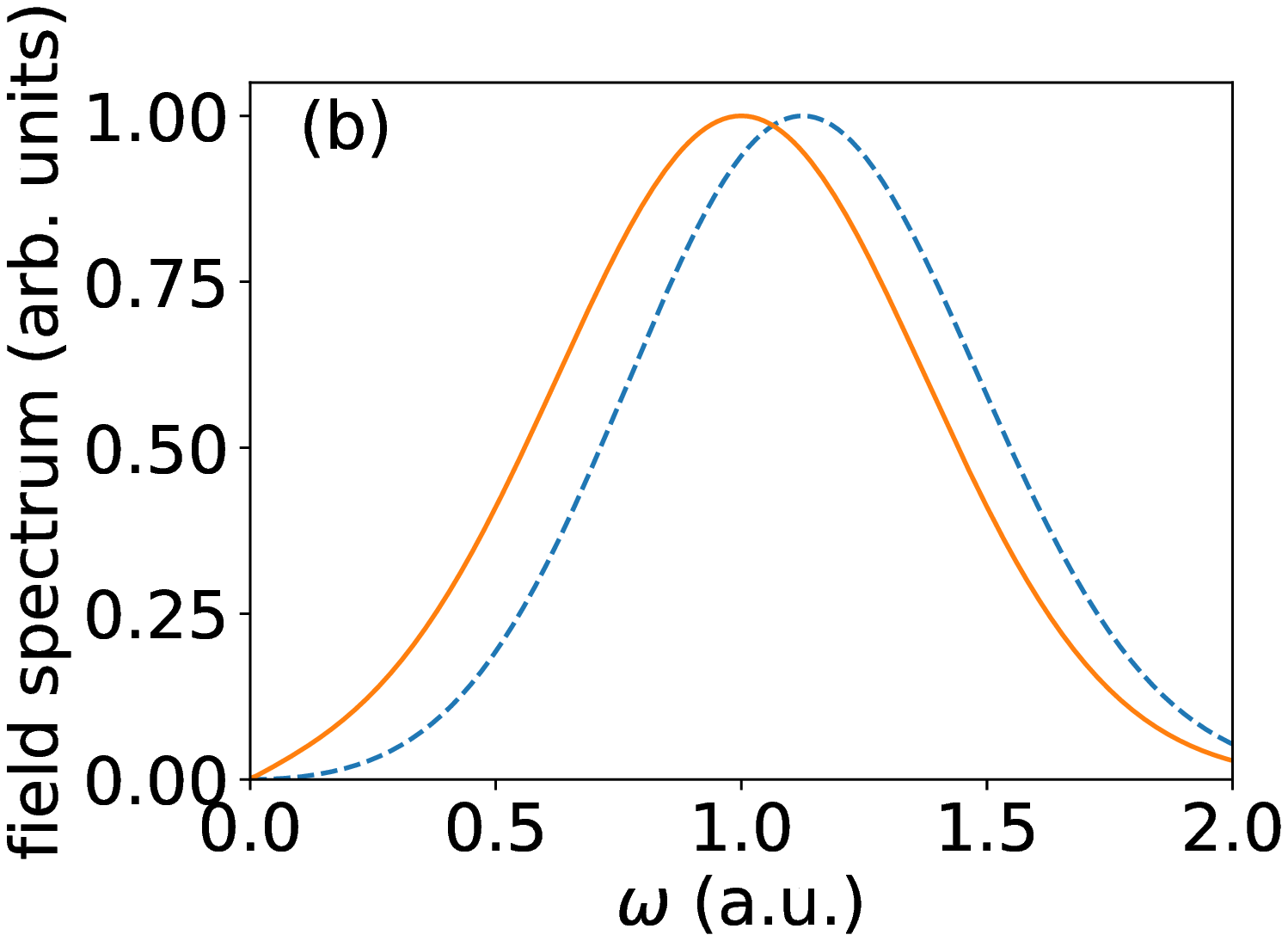}
   \includegraphics[width=0.49\linewidth]{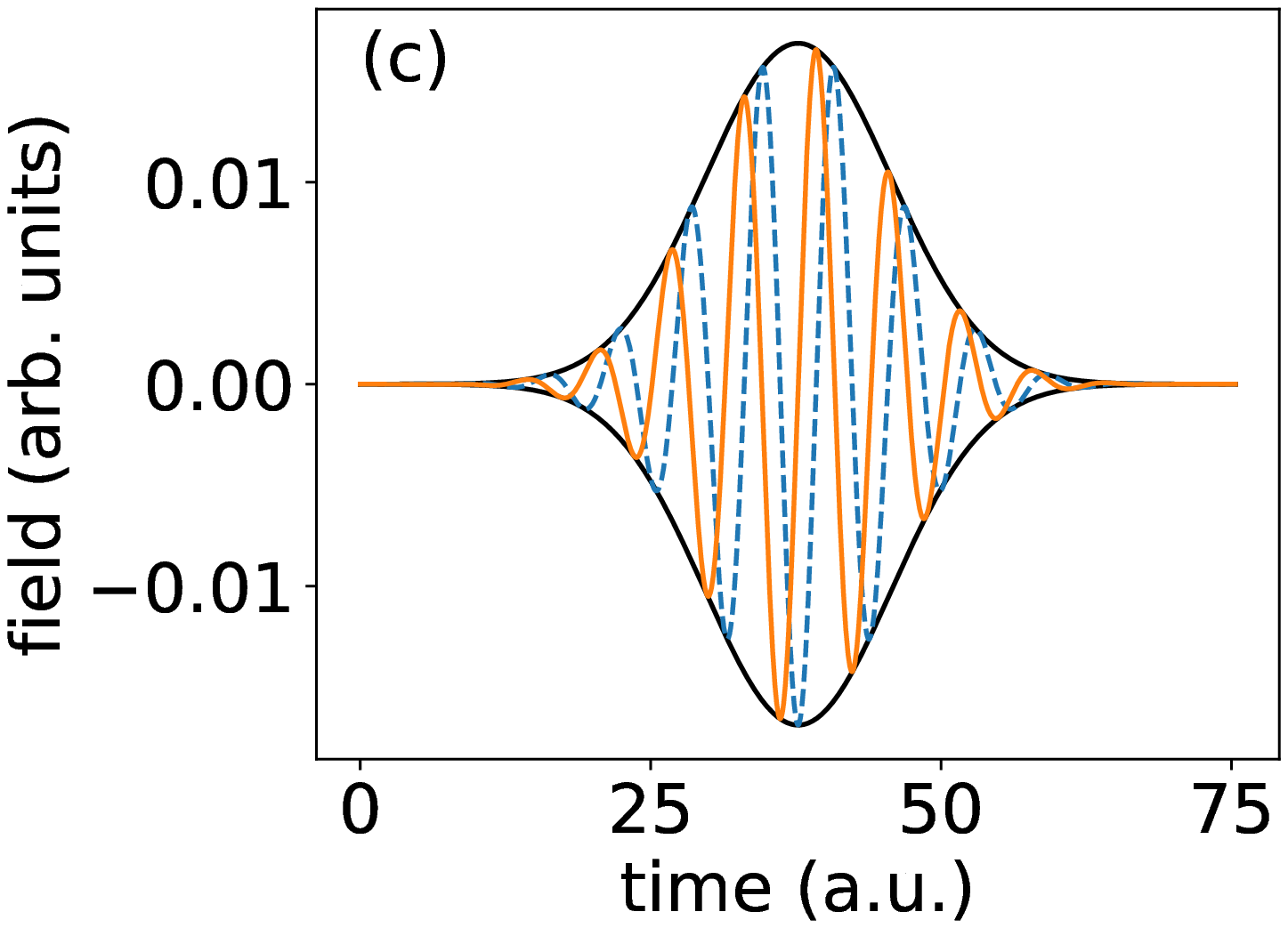}
   \includegraphics[width=0.49\linewidth]{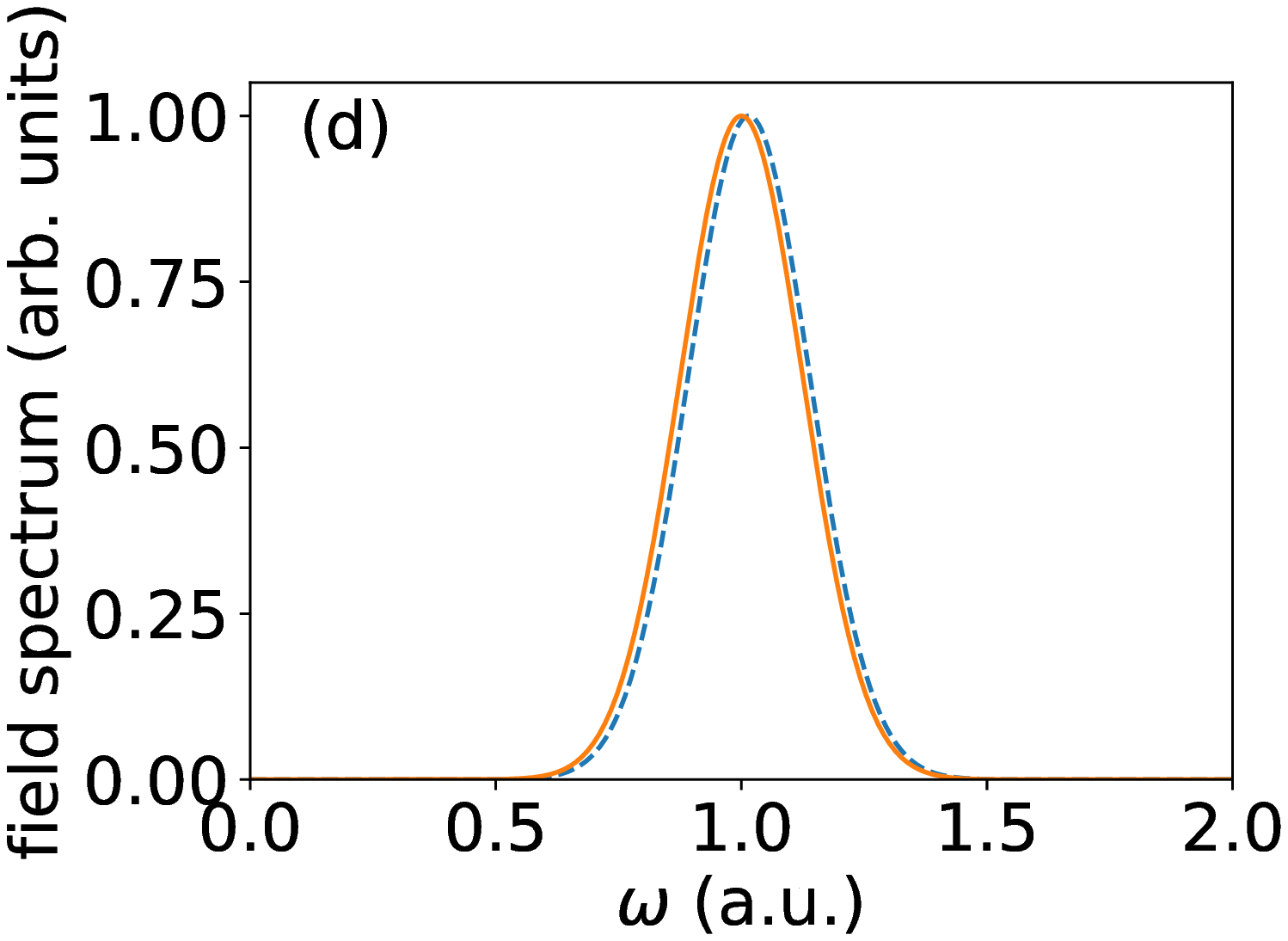}
   
\caption{(Color online)
Temporal (left) and spectral (right) distributions of vector potential (solid lines) and electric field (dashed line) for pulses with FWHM of 1 cycle (top), and 3 cycles (bottom) at central frequency $\omega_A = 1.0$ a.u.\  of the vector potential. Also shown is the Gaussian envelope of the vector potential. 
} 
  \label{fig:spectra}
\end{figure}

The frequency shift due to the difference in central frequencies is illustrated in Fig.~\ref{fig:spectra}, where the vector potential $A$ and the electric field $E$, obtained from Eqs. (\ref{eq:vectorpotential}) and (\ref{eq:efield}) for a Gaussian envelope $f(t)$, are compared in both the time and frequency domain. While the temporal behavior is satisfactory, the spectral distributions reveal different central frequencies. We define the central frequency $\omega_A$ ($\omega_E$) as the location of the maximum in the spectral density $|\tilde{A}(\omega)|$ ($|\tilde{E}(\omega)|$). The discrepancy is much greater for the 1-cycle full-width at half-max (FWHM) pulse (top), than for the 3-cycle FWHM pulse (bottom).

\begin{table}
\begin{center}
 \begin{tabular}{||c | c | c | c||} 
 \hline
 Name & Envelope Function $f(t)$ & $ N/N_{\text{FWHM}}$ & $\gamma_2$ \\ [0.5ex] 
 \hline\hline
 Gaussian & $e^{-(t/T)^2}$ & 0.849 & 0 \\ 
 \hline
 Cos$^2$ & $\begin{cases} 
      \cos^2\left(\frac{t}{T}\right) & -\frac{\pi}{2} \leq \frac{t}{T} \leq \frac{\pi}{2} \\
      0 & \text{otherwise}
   \end{cases}$ & 0.723 & -0.594 \\
 \hline
  Cos$^4$ &  $\begin{cases} 
       \cos^4\left(\frac{t}{T}\right) & -\frac{\pi}{2} \leq \frac{t}{T} \leq \frac{\pi}{2} \\
       0 & \text{otherwise}
    \end{cases}$ & 0.777 & -0.381 \\
 \hline
 Sech & $\text{sech}(\frac{t}{T})$ & 1.19 & 2.00 \\
 \hline
\end{tabular}
\end{center}
 \caption{Several common analytic pulse envelopes. The conversion factor between $N$, as defined in Eq.~(\ref{eq:nbar}) (number of cycles within one standard deviation to either side of the maximum), and the more typical $N_{\text{FWHM}}$ (number of cycles in the full-width half-maximum of the electric field) is given. The last column shows the excess kurtosis $\gamma_2$, defined in Eq.~(\ref{eq:gamma2}), which is independent of the pulse duration $T$.}
 \label{tab:envelopes}
\end{table}

For the further analysis, we note that in order for the central frequencies to be well-defined and consistent with the definition of $\omega_A$ in Eq. (\ref{eq:vectorpotential}), we make several assumptions about the envelope $f(t)$:
\begin{itemize}
  \item $f(t)$ is nonnegative and continuously differentiable,
  \item $f(t)$ falls off at least exponentially for large $|t|$, 
\item $f(t)$ contains no appreciable Fourier components larger than $\omega_A$.
\end{itemize}
These assumptions could be relaxed significantly, but they are sufficient for the present discussion and all practical purposes. The ratio of the central frequencies is given by the leading terms of an asymptotic expansion in $1/N$ as (see Appendix \ref{sec:appendix-frequency}):
\begin{eqnarray}
\label{eq:fshift}
\frac{\omega_E}{\omega_A} 
&=& 
\frac{1+\sqrt{1+4(\pi N)^{-2}}}{2}+ \frac{\gamma_2}{6\pi^4} N^{-4} + O( N^{-6})
\\
\label{eq:fshiftapprox}
&\approx& 
\frac{1+\sqrt{1+4(\pi N)^{-2}}}{2}.
\end{eqnarray}
Here
\begin{equation}
\begin{split}
\label{eq:nbar}
 N \equiv \frac{\omega_A}{\pi} \sqrt{\frac{\int_{-\infty}^{\infty} (t-t_0)^2 f(t) dt}{\int_{-\infty}^{\infty} f(t) dt}}.
\end{split}
\end{equation}
is the number of cycles within one standard deviation to either side of the pulse center, with
\begin{equation}
\begin{split}
t_0 \equiv \frac{\int_{-\infty}^{\infty} t f(t) dt}{\int_{-\infty}^{\infty} f(t) dt},
\end{split}
\end{equation}
and lastly $\gamma_2$ is the excess kurtosis of the envelope
\begin{equation}
\label{eq:gamma2}
\gamma_2 \equiv \frac{\left[\int_{-\infty}^{\infty} (t-t_0)^4 f(t) dt\right]\left[\int_{-\infty}^{\infty} f(t) dt\right]}{\left[\int_{-\infty}^{\infty} (t-t_0)^2 f(t) dt\right]^2}-3.
\end{equation}
Note that $N$ is proportional to the more typical $N_{\text{FWHM}}$ (number of cycles in the FWHM of $f(t)$), but the ratio $N/N_{FWHM}$ depends on the shape of the envelope (c.f., Table \ref{tab:envelopes}). 

\begin{figure}[t]
\centering
   \includegraphics[width=0.8\linewidth]{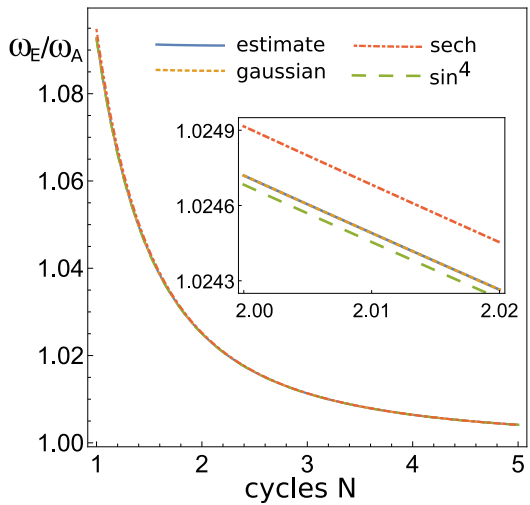}
\caption{(Color online)
Ratio $\omega_E/\omega_A$ as a function of the normalized number of cycles $ N$ defined in Eq.~(\ref{eq:nbar}). The numerical results were calculated by maximizing $\tilde{E}(\omega)$ for Gaussian (dotted line), $\text{sech}$ (dashed-dotted line), and $\sin^4$ (dashed line) envelopes and are compared with the simple analytic estimate (solid line) given in Eq.~(\ref{eq:fshiftapprox}). The inset reveals a slight dependence on envelope shape, which can be attributed to the correction term in Eq.~(\ref{eq:fshift}).
}
  \label{fig:shift}
\end{figure}

Eq.~(\ref{eq:fshift}) indicates that the ratio $\omega_E/\omega_A$ depends on the number of cycles and the pulse shape, but it is independent of peak intensity, carrier envelope phase, ellipticity, and $\omega_A$ itself. In fact, the first term of the expansion, Eq.~(\ref{eq:fshiftapprox}), is a very accurate estimate even for single cycle pulses, showing that the shift is nearly independent of pulse shape. This can be seen from the comparison between the predictions based on Eq.~(\ref{eq:fshiftapprox}) with the exact frequency shift for a variety of pulse shapes in Fig.~\ref{fig:shift}. The exact shift was calculated by numerically maximizing the analytic expressions for $|\tilde{E}(\omega)|$. The slight dependence on pulse shape is visible in the inset; the differences are described well by the correction term in Eq.~(\ref{eq:fshift}), which involves the excess kurtosis $\gamma_2$.

In the next section, we demonstrate that $\omega_E$ is the physically observable and relevant central frequency. Therefore, when modeling the interaction with a pulse using Eqs.~(\ref{eq:vectorpotential}) and (\ref{eq:efield}), one should determine $\omega_A$  such that it corresponds to the correct $\omega_E$. There are two different methods to do this. The first one is to specify $N$, and use Eq.~(\ref{eq:fshiftapprox}) to obtain $\omega_A$. However since $N$ depends implicitly on $\omega_A$ through Eq.~(\ref{eq:nbar}), the envelope $f(t)$ must be stretched in time by the same factor such that $N$ remains unchanged. The second method is to specify $f(t)$ instead of $N$. In that case, substituting Eq.~(\ref{eq:nbar}) into Eq.~(\ref{eq:fshiftapprox}) and solving for $\omega_A$ yields
\begin{equation}
\label{eq:alternative}
\omega_A \approx \omega_E - \frac{\int_{-\infty}^{\infty} f(t) dt}{\omega_E\int_{-\infty}^{\infty} (t-t_0)^2 f(t) dt}.
\end{equation}
Greater accuracy could be obtained in either case by including the correction term in Eq. (\ref{eq:fshift}); however, the results in Fig.~\ref{fig:shift} show that this is in general not necessary. 

\section{Applications}
\label{sec:applications}

In this section we present results of numerical calculations which exemplify effects of the frequency shift on observables related to excitation, ionization and high harmonic generation induced by short laser pulses. To this end, we solved the 3D one-electron time-dependent Schr\"odinger equation (TDSE) in velocity gauge:
\begin{equation}
i\frac{\partial}{\partial t}\psi(\mathbf{r},t) = \left[\frac{\mathbf{p}^2}{2} - \frac{\mathbf{A}(t) \cdot \mathbf{p}}{c} + V(\mathbf{r})\right]\psi(\mathbf{r},t)
\end{equation}
and length gauge
\begin{equation}
i\frac{\partial}{\partial t}\psi(\mathbf{r},t) = \left[\frac{\mathbf{p}^2}{2} + \mathbf{E}(t) \cdot \mathbf{\mathbf{r}} + V(\mathbf{r})\right]\psi(\mathbf{r},t)
\end{equation}
for atomic hydrogen with a soft-core Coulomb potential
\begin{equation}
V(\mathbf{r}) = -\frac{1}{\sqrt{\mathbf{r}^2+\alpha ^2}}.
\end{equation}
We consider a linearly polarized laser pulse within the dipole approximation, so $\mathbf{A}(t)=A(t)\hat{z}$ and $\mathbf{E}(t)=E(t)\hat{z}$. Taking advantage of azimuthal symmetry, the wavefunction can be represented in 2D cylindrical coordinates $\rho$ and $z$. We used the second order finite difference method for spatial derivatives and the fully implicit second order Crank-Nicholson method for time propagation (for more details on the numerical implementation, see \cite{venzke-submitted}). The laser field magnitudes $A(t)$ and $E(t)$ were defined as in Eq.~(\ref{eq:vectorpotential}) and Eq.~(\ref{eq:efield}), with a Gaussian envelope function $f(t)$. The results presented below were obtained in velocity gauge, but additional test calculations in length gauge have confirmed that the results are gauge invariant.

For the single photon ionization (Sec.~\ref{sec:single_photon}) calculations we used $\alpha = 0$, giving a ground state energy of $E_{1s} = -0.5109$ a.u. To ensure the wavefunction remains on the grid for our calculation of the photoelectron spectrum, the grid extended 500 a.u.\  in the $\rho$-direction and 1000 a.u in the $z$-direction, with an exterior complex scaling absorbing boundary in the outer 50 a.u. A grid spacing of 0.2 a.u.\  and a time step of 0.1 a.u were used.

For our studies of excitation (Sec.~\ref{sec:excitation}) and high harmonic generation (Sec.~\ref{sec:HHG}) we used $\alpha=0.029$ a.u., giving a ground state energy of $E_{1s} = -0.5001$ a.u.\  and an excited state energy of $E_{2p} = -0.12504$ a.u.\  In this case the grid extended over 100 a.u.\  in the $\rho$-direction and 200 a.u.\ in the $z$-direction, with an absorbing boundary over the outer 5 a.u.  A grid spacing of 0.1 a.u.\  and a time step of 0.1 a.u were used.

\subsection{Single Photon Ionization}
\label{sec:single_photon}


First, we consider single photon ionization of the hydrogen atom by a few-cycle laser pulse with peak intensity $10^{13}$ W/cm$^2$ and central frequency $\omega_{\text{central}} = 2.0$ a.u. The central frequency is implemented either by setting $\omega_A=\omega_{\text{central}}$ or setting $\omega_E=\omega_{\text{central}}$, using the method described in the previous section. Photoelectron momentum spectra $P(k)$ were obtained by the following procedure: the TDSE was propagated for five times the FWHM pulse duration plus an additional 100 a.u.\  in time, then all bound states with principle quantum number $n\le 8$ were projected out, and lastly the remaining unbound wavepacket $\psi_{\text{ionized}}$ was projected onto spherical waves up to $l_{\text{max}}=5$. That is,
\begin{equation}
P(k) = \sum_{l=0}^{l_{\text{max}}} \left|\int j_{l}(kr)Y^*_{l0}(\hat{r})\psi_{\text{ionized}}(\vec{r})d^{3}\vec{r}~\right|^2 ~.
\end{equation}

\begin{figure}[t]
\centering
   \includegraphics[width=0.75\linewidth]{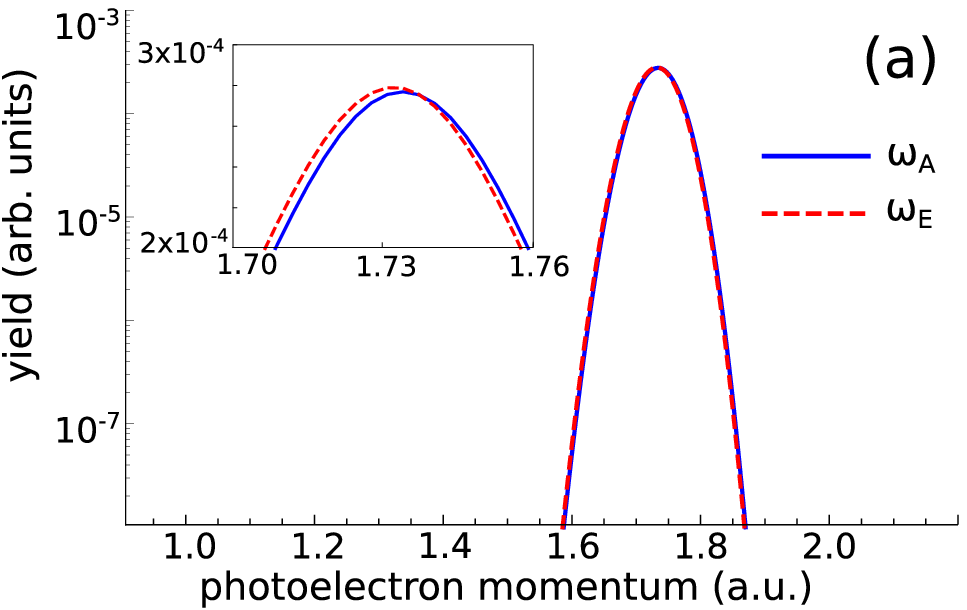}
   \includegraphics[width=0.75\linewidth]{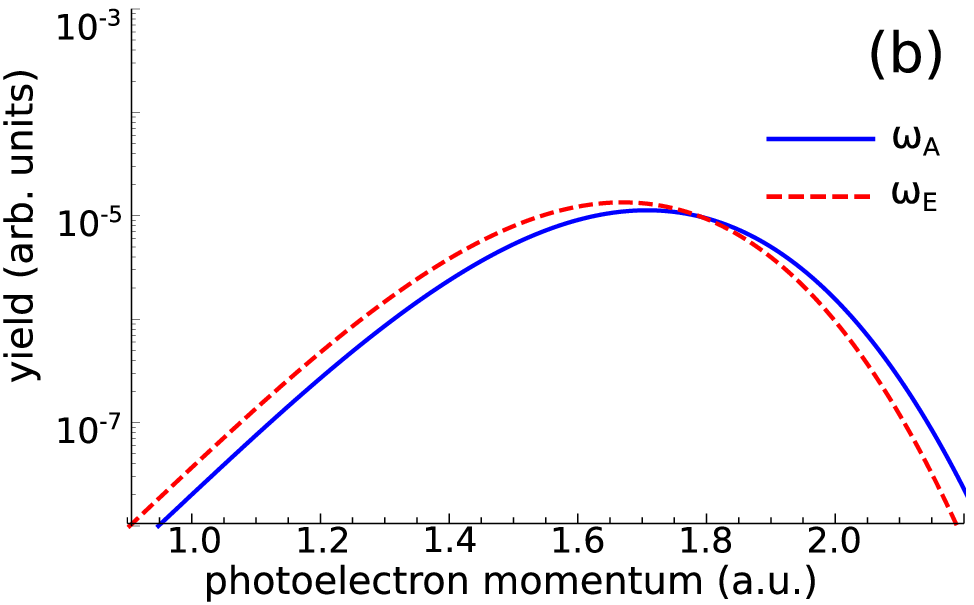}
\caption{(Color online)
Photoelectron spectra $P(k)$ as function of photoelectron momentum obtained for interaction of hydrogen atom with laser pulses at central frequencies $\omega_A=2$ (dashed line) and $\omega_E=2$ (solid line) and duration of (a) 10 cycles and (b) 2 cycles FWHM.
}
  \label{fig:ionizationA}
\end{figure}

The results in Fig.~\ref{fig:ionizationA} show that in fact the photoelectron spectra for central frequency $\omega_A = 2$ a.u.\  (solid lines) and $\omega_E = 2$ a.u.\  (dashed lines) do not agree due to the frequency shift. As expected, the discrepancy is larger for 2 cycle FWHM (panel (b)) than for 10 cycle FWHM (panel (a)) pulses. These results however raise the question whether the central frequency of the vector potential $\omega_A$ or the central frequency of the electric field $\omega_E$ is the relevant quantity for further physical interpretation or a comparison with experimental data. To address this question, we consider the resonant population transfer between bound states in the next subsection.

\subsection{Excitation}
\label{sec:excitation}

Next, we examine transitions to the $n=2$ orbitals in the hydrogen atom as a function of both pulse length and central frequency of the vector potential ($\omega_A$) and the electric field ($\omega_E$).
Typically, the excitation probability is greatest when the central frequency of the laser matches the resonant frequency for $n_p$ photon absorption, given by
\begin{equation}
\omega_{\text{res}} = |E_\text{final} - E_\text{initial}|/n_p=0.375/n_p.
\end{equation}
In view of the predicted frequency shift between the central frequencies $\omega_A$ and $\omega_E$, we therefore expect that the results for resonant excitation will provide insights into the physical relevance of $\omega_A$ vs.~$\omega_E$.

\begin{figure}[t]
\centering
   \includegraphics[width=\linewidth]{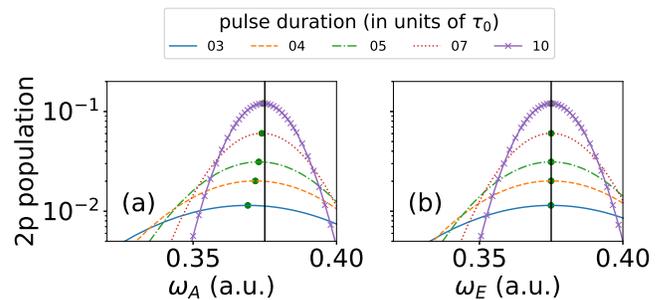}
\caption{(Color online) 
Population in $2p$ state following one-photon excitation of hydrogen atom with a laser pulse as a function of $\omega_A$ (left) and $\omega_E$ (right) for different pulse lengths at peak intensity $10^{12}$ W/cm$^2$. Each line represents results obtained for a fixed pulse duration in terms of $\tau_0=405$ as. The vertical line marks the energy difference between $2p$ and initial $1s$ state. The green dots indicate the maximum excited state population for each pulse duration. 
}
  \label{fig:h-excited_2p}
\end{figure}

\begin{figure}[t]
\centering
      \includegraphics[width=\linewidth]{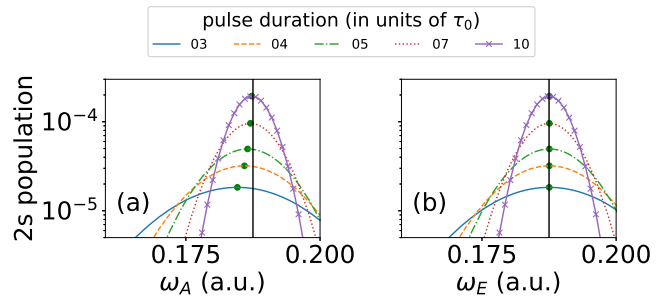}
\caption{(Color online)
Same as Fig.~\ref{fig:h-excited_2p} for population in $2s$ state following 2-photon excitation with $\tau_0=811$ as. The vertical line marks half of the energy gap between $1s$ and $2s$ representing the resonance condition for the two photon process.
}
  \label{fig:h-excited_2s}
\end{figure}

In the corresponding calculations the peak laser intensity was chosen as $10^{12}$ W/cm$^2$ to ensure significant population transfers for short pulses. Despite the moderate intensity, we may neglect the effect of the field on the atomic energy levels in our analysis. In Fig.~\ref{fig:h-excited_2p} the population in the $2p$ state due to single photon excitation is presented as function of $\omega_A$ (Fig.~\ref{fig:h-excited_2p}a) and $\omega_E$ (Fig.~\ref{fig:h-excited_2p}b) for various pulse lengths. For long pulses the peak in the population (marked by a green dot) occurs at the expected frequency $\omega_\text{res}$ (marked by vertical line) for a resonant transition in both distributions. When the pulse length is decreased, the peak in the distribution as a function of $\omega_E$ (Fig.~\ref{fig:h-excited_2p}b) remains at $\omega_\text{res}$. In contrast, the peak shifts significantly towards lower frequencies in the distribution as a function of $\omega_A$ (Fig.~\ref{fig:h-excited_2p}a) due to the frequency shift. Therefore, the central frequency of the electric field is the physically relevant quantity for interpreting laser induced excitation processes.

These conclusions are further supported by the results for two-photon excitation from the $1s$ to the $2s$ state in Fig.~\ref{fig:h-excited_2s}. Whereas the peak of the population as a function of $\omega_E$ (panel b) occurs at $\omega_\text{res}$, independent of the pulse duration, the peak of the population as a function of  $\omega_A$ once again shifts to lower frequencies in Fig.~\ref{fig:h-excited_2s}(a). We note that, if the population as function of $\omega_A$ in Figs.~\ref{fig:h-excited_2p} and \ref{fig:h-excited_2s} were used to determine the energy difference $|E_\text{final} - E_\text{initial}|$, the error caused by the frequency shift would be twice as much in the two-photon case as in the one-photon case, accounting for the difference in $\tau_0$. This indicates that multiphoton processes may be affected by the frequency shift even more than few-photon processes. To further illustrate this point, we examine high harmonic generation in subsection~\ref{sec:HHG}.

\subsection{High Harmonic Generation}
\label{sec:HHG}

Finally, we consider a highly nonlinear laser induced process. High harmonic generation (HHG) in atoms can be described as absorption of an odd number of photons leading to the excitation of a electron, followed by the emission of a single photon as the electron recombines into the ground state. Based on the results above, we expect that in this nonlinear process the frequency shift $\Delta\omega$ between $\omega_A$ and $\omega_E$ will lead to a shift of the energy of the $n_p$th harmonic by $n_p\Delta\omega$. In our calculations the HHG spectrum has been obtained by a Fourier transformation of the time dependent dipole acceleration along the laser polarization direction. A Hanning filter was used to return the dipole acceleration to zero at the beginning and end of the simulation. 

\begin{figure}[t]
\centering
   \includegraphics[width=\linewidth]{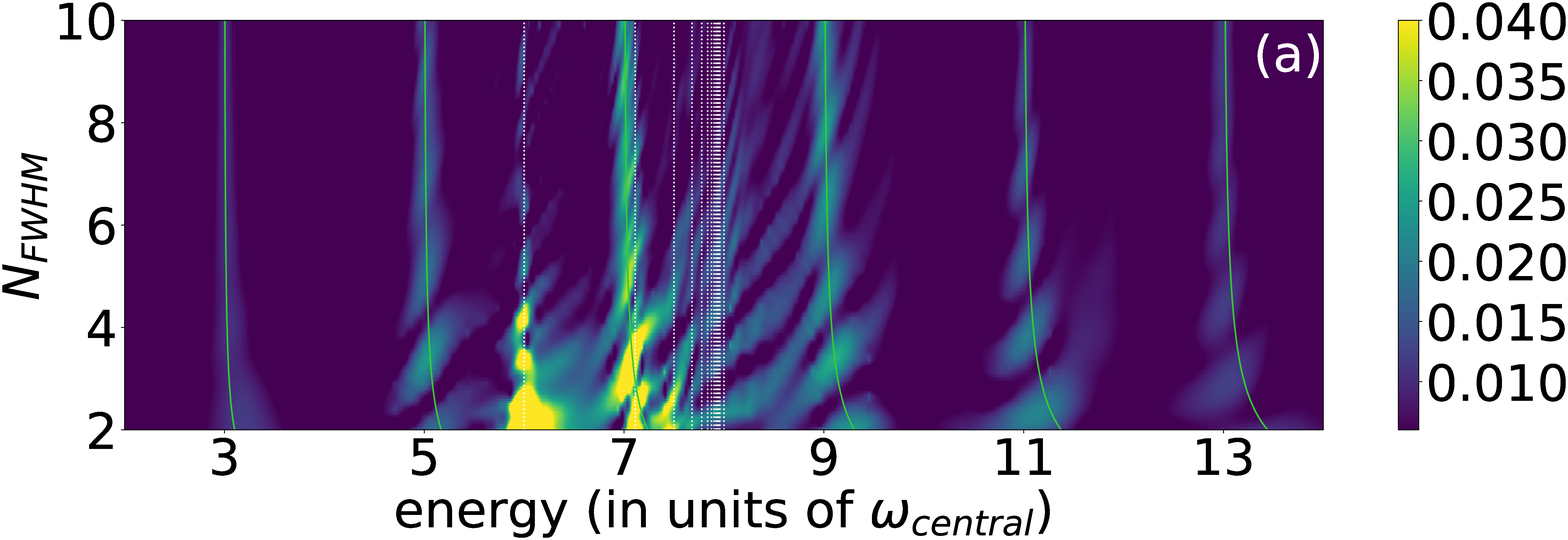}   		
   \includegraphics[width=\linewidth]{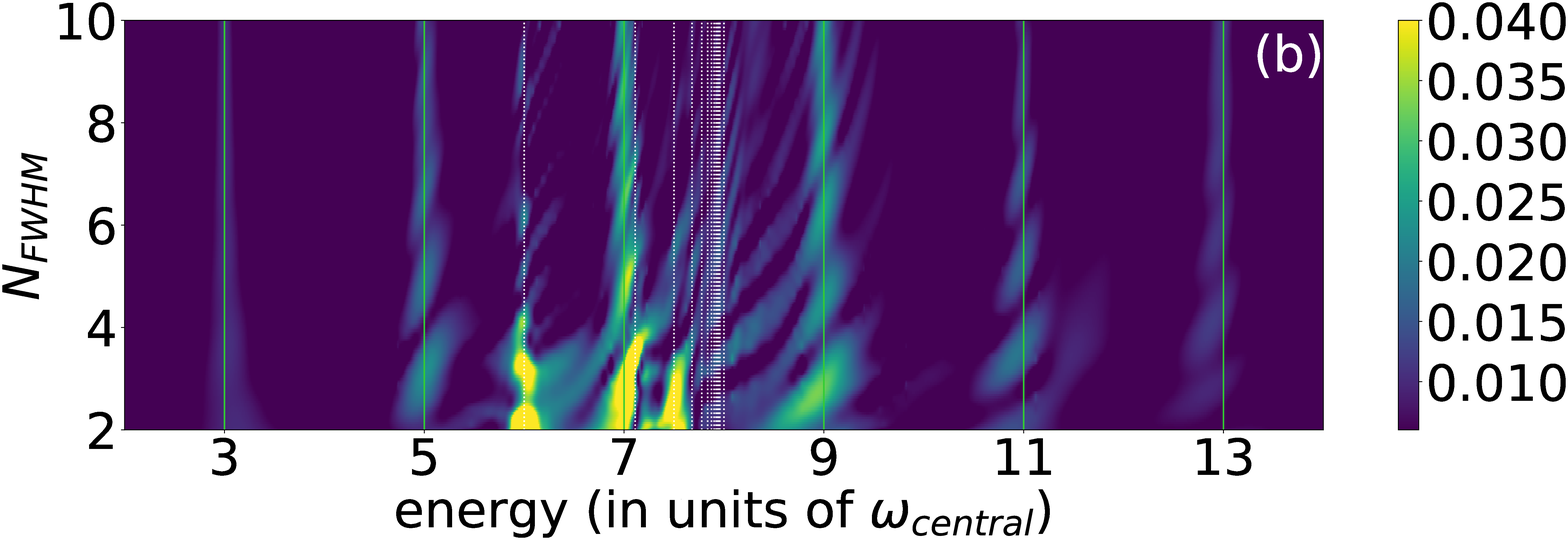}
\caption{(Color online)
HHG spectrum at driver wavelength 730 nm ($\omega_\text{central}$ = 0.0625 a.u.) vs.\ number of cycles $N_\text{FWHM}$. In the upper plot the central frequency $\omega_A = \omega_\text{central}$ while in the lower panel $\omega_E = \omega_\text{central}$. The vertical white dashed lines mark field-free transition energies between excited states and the ground state, while the green solid lines mark the harmonic energies $n_p\omega_E$ with respect to the central frequency of the electric field.
}
  \label{fig:HHG}
\end{figure}

Fig.~\ref{fig:HHG} shows the various harmonics in a HHG spectrum as a function of the number of cycles in the driving laser pulse at a peak intensity of $1\times 10^{14}$ W/cm$^2$ and central frequencies $\omega_A=0.0625$ a.u.\ \ (upper panel) and $\omega_E=0.0625$ a.u.\ \ (lower panel), corresponding to a wavelength of 730 nm. The spectrum consists of odd harmonics and additional emission lines due to the population of excited states during the interaction with the laser pulse. While we will focus on the generation of harmonics, we note that the emission lines occur at photon energies between the 5th and 9th harmonics. The corresponding field-free energy differences between the excited states and the ground state in our numerical model of the hydrogen atom are marked, as reference, by white vertical dashed lines.

In the spectrum as function of multiples of $\omega_A$ (panel a) one can see that the centers of the harmonics do shift to energies larger than $n_p\omega_A$, as the pulse duration decreases. In fact, the energies of the harmonics follow the analytical predictions for $n_p\omega_E$ (green solid lines). As expected, the shift is as larger as larger the harmonic number. In contrast, in the HHG spectrum obtained as multiples of $\omega_E$ (panel b) the centers of the harmonics remain at the same energy, i.e.~$n_p\omega_E$ (green solid lines), as the pulse duration decreases. This confirms the importance of the shift between the central frequencies of the vector potential and the electric field in nonlinear processes driven by ultrashort pulses. Furthermore, the HHG results confirm that the central frequency of the electric field $\omega_E$ is the physical relevant quantity for the interpretation of light induced processes. Consequently, if in a numerical simulation or theoretical analysis the vector potential is set via Eq.~(\ref{eq:vectorpotential}) it is necessary to consider the frequency shift between $\omega_A$ and $\omega_E$ to avoid a misinterpretation of the results. Our analytical estimates of the frequency shift in Eq.~(\ref{eq:fshiftapprox}) and Eq.~(\ref{eq:alternative}) provide formulas to obtain $\omega_A$ from the physically relevant $\omega_E$.

\section{Summary}

We have shown that the definition of the electric field of a laser pulse via the derivative of the vector potential, which guarantees that both quantities vanish at the beginning and end of the pulse, implies that the central frequencies of the spectral distributions of the vector potential and electric field do not coincide. In our analysis we have derived an analytical estimate of the frequency shift, which shows that the shift mainly depends on the number of cycles in the pulse and becomes most relevant for few-cycle pulses. Utilizing results of numerical simulations we have analyzed how the frequency shift affects excitation, ionization and high harmonic generation induced by short laser pulses. The effect is found to be most noticeable in nonlinear strong-field processes since the frequency shift scales with the number of photons involved. Overall, the numerical results confirm that the central frequency of the electric field is the physically relevant quantity for the interpretation of the light induced processes. Thus, the shift should be taken into account when setting the central frequency of the vector potential in numerical simulations to avoid potential misinterpretation of the theoretical results, specifically when compared to experimental data.

\section*{Acknowledgements}

J.V. and A.B. were supported via a grant from the U.S.\ Department of Energy, Division of Chemical Sciences, Atomic, Molecular and Optical Sciences Program, while T.J., Z.X. and A.J.-B.\ were supported by a grant from the U.S.\ National Science Foundation (Award Nos. PHY-734006). This work utilized the RMACC Summit supercomputer, which is supported by the National Science Foundation (Award Nos. ACI-1532235 and ACI-1532236), the University of Colorado Boulder, and Colorado State University. The Summit  supercomputer is a joint effort of the University of Colorado Boulder and Colorado State University.

\appendix

\section{Derivation of frequency shift}
\label{sec:appendix-frequency}

In this Appendix we derive the expansion of the frequency shift in Eq.~(\ref{eq:fshift}). To do this, we first introduce a fixed point iteration that will be used to calculate the leading terms of an asymptotic expansion in $1/N$. Here the limit $N\rightarrow\infty$ refers to keeping $\omega_A$ fixed, but scaling the envelope $f(t)\rightarrow f(\lambda t)$ so that $N\rightarrow \lambda^{-1}N$ according to Eq. (\ref{eq:nbar}), and taking the limit $\lambda\rightarrow 0$.

We start with some definitions. Fourier transforms will be denoted by a tilde, so that for any function $g(t)$,
\begin{equation}
\begin{split}
\label{eq:fourierTransform}
\tilde{g}(\omega) &\equiv \frac{1}{2\pi}\int_{-\infty}^{\infty}g(t)e^{-i\omega t} dt .
\end{split}
\end{equation}
The vector potential of Eq.~(\ref{eq:vectorpotential}) can be written in the frequency domain as
\begin{equation}
\begin{split}
\label{eq:afieldft}
\tilde{A}(\omega) &= \frac{1}{2}e^{i\phi}\tilde{f}(\omega-\omega_A) + \frac{1}{2}e^{-i\phi}\tilde{f}(\omega+\omega_A).
\end{split}
\end{equation}
The assumptions on $f(t)$ made in Sec. \ref{sec:pulse} imply that the second term can be neglected for $\omega>0$.
And since $f(t)$ is nonnegative, the spectral distribution $|\tilde{A}(\omega)|$ peaks at $\omega_A$.

In the frequency domain, $\tilde{A}$ and $\tilde{E}$ are related by
\begin{equation}
\begin{split}
\label{eq:efieldft}
\tilde{E}(\omega) &= -\frac{i\omega}{c}\tilde{A}(\omega).
\end{split}
\end{equation}
The factor of $\omega$ shifts the peak so that $\omega_E\geq\omega_A$.

\begin{theorem}
For sufficiently large $N$, the ratio of the central frequencies is $\omega_E/\omega_A = 1+X$, where $X$ is a fixed point of the following iteration:
\begin{equation}
\begin{split}
\label{eq:fixedPoint}
&x_{i+1} - x_{i} = \frac{1}{\pi^2 N^2}\left[1+ \omega_A (1+x_i)g(x_i) \right]\\
&\text{with} \quad g(x) \equiv \frac{d}{d\omega}\left[ \log|\tilde{f}(\omega)|\right]\Big|_{\omega = \omega_A x}
\end{split}
\end{equation}
and $N$ is defined in Eq.~(\ref{eq:nbar}).
\end{theorem}
\textit{Proof:} Using Eqs.~(\ref{eq:afieldft}) and  (\ref{eq:efieldft}),
\begin{equation}
\begin{split}
\tilde{E}(\omega) &= -\frac{i\omega}{2c}e^{i\phi}\tilde{f}(\omega-\omega_A) - \frac{i\omega}{2c}e^{-i\phi}\tilde{f}(\omega+\omega_A).
\end{split}
\end{equation}
For $\omega>0$ the second term can be dropped. Therefore, $\omega_E$, which is defined as the position of the maximum of $|\tilde{E}(\omega)|$, also maximizes
\begin{equation}
\begin{split}
\label{eq:toMax}
&\log\left[\omega|\tilde{f}(\omega-\omega_A)|\right].
\end{split}
\end{equation}
To locate the maximum, we set the derivative equal to zero, substitute $X=\omega_E/\omega_A - 1$ and have
\begin{equation}
\begin{split}
\label{eq:toMax2}
0&=\omega_E^{-1} + \frac{d}{d\omega}\left[ \log|\tilde{f}(\omega-\omega_A)|\right]\Big|_{\omega = \omega_E}\\
0&=1 + \omega_E\frac{d}{d\omega'}\left[ \log|\tilde{f}(\omega')|\right]\Big|_{\omega' = \omega_AX}\\
0&=1 + \omega_A(1+X)g(X) .
\end{split}
\end{equation}
Therefore, $X$ is a fixed point of the iteration. $\square$

Standard techniques can be applied to analyze this fixed point iteration (see e.g., \cite{atkinson89}). We therefore state the following theorem 2 without proof:
\begin{theorem}
Starting from $x_1 = 0$, and assuming $N$ is not too small, the fixed point iteration defined in Eq. (\ref{eq:fixedPoint}) converges to the smallest positive fixed point, which is $X=\omega_E/\omega_A - 1$. The rate of convergence is $O(N^{-2})$, and $X$ itself is also $O(N^{-2})$.
\end{theorem}

Before proceeding, we note that the fix point iteration provides an algorithm to compute the frequency shift numerically. In the remainder of the appendix we will however use the two theorems above to derive an asymptotic expansion in the limit of large $N$, which gives us the expression in Eq. (\ref{eq:fshift}). To this end, we apply the iteration to a formal truncated power series in $1/N$. Because of the $O(N^{-2})$ convergence, each iteration gives an additional term in the asymptotic expansion. We compute only the first two terms here, leaving the error at $O(N^{-6})$.

To do so, we would like to expand $g(x)$ in a Taylor series about $x=0$.
The logarithm of a Fourier transform resembles the cumulant generating function in statistics \cite{kendall87},
\begin{equation}
\label{eq:logExpansion}
\log\left[\tilde{f}(\omega)\right] = \log\left(\frac{\kappa_0}{2\pi}\right)+\sum_{n=1}^{\infty}\kappa_n \frac{(i\omega)^n}{n!}.
\end{equation}
Accounting for the fact that $f(t)$ is not normalized, the first few cumulants are defined as
\begin{equation}
\begin{split}
\label{eq:cumulants}
\kappa_0 &= \int_{-\infty}^{\infty} f(t)dt \\
\kappa_1 &= \kappa_0^{-1}\int_{-\infty}^{\infty} t f(t)dt \\
\kappa_2 &= \kappa_0^{-1}\int_{-\infty}^{\infty}
(t-\kappa_1)^2 f(t)dt \\
\kappa_3 &= \kappa_0^{-1}\int_{-\infty}^{\infty}
(t-\kappa_1)^3 f(t)dt \\
\kappa_4 &= \kappa_0^{-1}\int_{-\infty}^{\infty}
(t-\kappa_1)^4 f(t)dt - 3\kappa_2^2\\
\kappa_5 &= \kappa_0^{-1}\int_{-\infty}^{\infty}
(t-\kappa_1)^5 f(t)dt - 10\kappa_3\kappa_2~.
\end{split}
\end{equation}
Substituting Eq. (\ref{eq:logExpansion}) into the definition of $g$ yields
\begin{equation}
\begin{split}
\label{eq:gTaylor}
g(x)=& \frac{d}{d\omega}\left[ \log|\tilde{f}(\omega)|\right]\Big|_{\omega = \omega_A x}\\
=&\frac{d}{d\omega}\text{Re}\left[\log \tilde{f}(\omega)\right]\Big|_{\omega = \omega_A x}\\
=&\frac{d}{d\omega}\left[\sum_{n=1}^{\infty}(-1)^{n}\kappa_{2n} \frac{\omega^{2n}}{(2n)!}   \right]\Big|_{\omega = \omega_A x}\\
=&\sum_{n=1}^{\infty}(-1)^{n}\kappa_{2n} \frac{(\omega_A x)^{2n-1}}{(2n-1)!} \\
=&-\kappa_2\omega_A x + \kappa_4 \frac{(\omega_A x)^{3}}{6} + O(x^5)
\end{split}
\end{equation}
Comparing Eq. (\ref{eq:cumulants}) to Eqs. (\ref{eq:nbar}) and (\ref{eq:gamma2}) indicates
\begin{equation}
\begin{split}
\label{eq:correspondence}
\kappa_2 = \left(\frac{\pi N}{\omega_A}\right)^2, \qquad\quad
\kappa_4 = \gamma_2 \left(\frac{\pi N}{\omega_A}\right)^4.
\end{split}
\end{equation}

Using the Taylor expansion for $g(x)$, we apply the fixed point iteration in Eq. (\ref{eq:fixedPoint}) to a formal power series truncated at order $O(N^{-6})$:
\begin{equation}
\begin{split}
\label{eq:iteration}
x_1 &= 0\\
x_2 &= (\pi N)^{-2}\\
x_3 &= 2(\pi N)^{-2} + \omega_A[(\pi N)^{-2}+(\pi N)^{-4}]g((\pi N)^{-2})\\
    &= (\pi N)^{-2}+(\tfrac{\gamma_2}{6}-1)(\pi N)^{-4} + O(N^{-6})\\
x_4 &= (\pi N)^{-2}+(\tfrac{\gamma_2}{6}-1)(\pi N)^{-4} + O(N^{-6})
\end{split}
\end{equation}
Since $x_3=x_4$ up to $O(N^{-6})$, the iteration has converged after two iterations to the leading asymptotic terms. Conveniently, all terms in the full asymptotic expansion contain only $\kappa_2$ but no higher cumulants and can be re-summed into a square root:
\begin{equation}
\label{eq:fshiftapp}
X = \frac{-1+\sqrt{1+4(\pi N)^{-2}}}{2}+ \frac{\gamma_2}{6\pi^4}N^{-4} + O(N^{-6})
\end{equation}
and hence,
\begin{equation}
\frac{\omega_E}{\omega_A}  = \frac{1+\sqrt{1+4(\pi N)^{-2}}}{2}+ \frac{\gamma_2}{6\pi^4}N^{-4} + O(N^{-6}).
\end{equation}
The fixed point iteration can be used in this way to calculate as many terms in the asymptotic expansion as desired. Only even powers of $N$ appear, and the coefficient of the $N^{-2n}$ term contains even cumulants of $f(t)$ up to $\kappa_{2n}$.

\end{document}